\documentclass[%
 reprint,
 amsmath,amssymb,
 aps,
]{revtex4-2}

\usepackage{graphicx}
\usepackage{dcolumn}
\usepackage{bm}
\usepackage{hyperref}

\begin{document}

\preprint{APS/123-QED}

\title{Flow Crossover and Parallel Outflow during Collisionless Magnetic Reconnection}

\author{Theerasarn Pianpanit}
\affiliation{
Department of Applied Radiation and Isotopes, Faculty of Science, Kasetsart University, Bangkok, Thailand
}

\author{Kittipat Malakit}
\email[Corresponding author, ]{kmalakit@gmail.com}
\affiliation{
Department of Physics, Faculty of Science and Technology, Thammasat University, Pathum Thani, Thailand
}

\author{Pakkapawn Prapan}
\affiliation{
Department of Physics, Faculty of Science, Mahidol University, Bangkok, Thailand
}

\author{Peera Pongkitiwanichakul}
\affiliation{
Department of Physics, Faculty of Science, Kasetsart University, Bangkok, Thailand
}

\author{Piyawat Suetrong}
\affiliation{
Department of Physics, Faculty of Science, Kasetsart University, Bangkok, Thailand
}

\author{Michael A. Shay}
\affiliation{
Department of Physics and Astronomy, University of Delaware, Newark, DE, United States
}

\author{Paul A. Cassak}
\affiliation{
Department of Physics and Astronomy and Center for KINETIC Plasma Physics, West Virginia University, Morgantown, WV, United States
}

\author{David Ruffolo}
\affiliation{
Department of Physics, Faculty of Science, Mahidol University, Bangkok, Thailand
}

\date{\today}

\begin{abstract}
Using particle-in-cell simulations that label ions and electrons according to their initial inflow region, we find that during 2D collisionless magnetic reconnection, the {\it bulk} flow of the plasma from each inflow side crosses paths with plasma from the other inflow side and crosses the midplane before being redirected into an outflow jet. This feature, which we term ``flow crossover,'' 
implies
mechanisms to generate bulk motion in a direction parallel to the magnetic field. We find that ions and electrons undergo different parallel driving mechanisms, leading to different flow crossover patterns.  The parallel bulk flow for ions is generated more locally within the ion diffusion region, whereas the parallel bulk flow for electrons is mostly generated outside the electron diffusion region. Consequently, the reconnection outflows are more of a parallel flow than a perpendicular flow, especially for the electron outflow. The flow crossover and the parallel outflow patterns occur not only in symmetric reconnection but also in the more complex scenario of a guide-field asymmetric reconnection, suggesting that it is a general feature of collisionless magnetic reconnection.
Because the plasma on one side of the outflow mostly originates from the inflow plasma on the other side, 
we predict that near an asymmetric reconnection site in a collisionless space plasma, {\it in situ} observations across the outflow region could reveal locally reversed gradients in plasma properties.
\end{abstract}

\maketitle



\begin{figure}[ht!]
\centering
\includegraphics[width=0.35\textwidth]{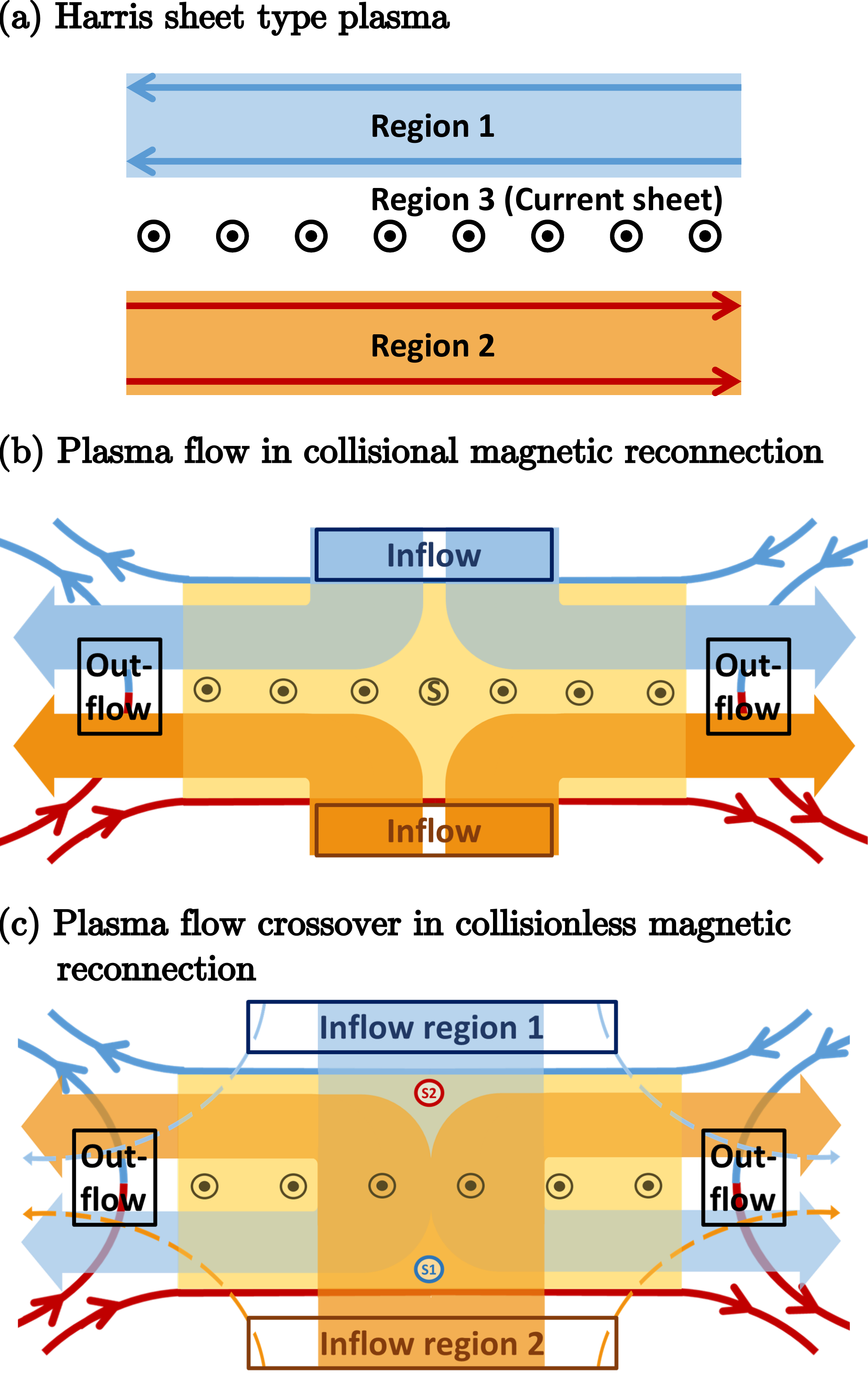}
\caption{(a) For a Harris sheet initial configuration, the plasma particles can be labeled by their initial region, where the inflow regions 1 and 2 are separated by the current sheet region 3. 
Dark-colored arrows indicate magnetic field lines. (b) Illustration of plasma flow (light-colored arrows) in collisional magnetic reconnection.  Yellow rectangle indicates the diffusion region, and ``S'' the stagnation point. (c) Illustration of our finding of plasma flow crossover in collisionless magnetic reconnection.  Now there are separate stagnation points for flow from each inflow region, labeled ``S1'' and ``S2,'' located opposite the midplane from the initial inflow region.
}
\label{fig:view}
\end{figure}

\textit{Introduction}---
Magnetic reconnection is a fundamental plasma process that underlies various astrophysical phenomena such as solar flares, coronal mass ejections, the Dungey cycles of planetary magnetospheres, and star formation \citep{yamada10}. During magnetic reconnection, inflowing plasmas bring 
magnetic field lines with an oppositely directed component
toward a diffusion region (\autoref{fig:view}(a)), where the frozen-in condition is broken.  Within this region there is a change in magnetic field line topology, which facilitates the conversion of energy stored in the magnetic field into thermal and bulk kinetic energy of the outflowing plasma. 
The physics of magnetic reconnection is somewhat different in collisional plasmas, i.e., when collisions play an important role in plasma dynamics, versus low collisionality plasmas that are relevant for reconnection in solar and stellar coronae, heliospheric phenomena, and planetary magnetospheres.


Influenced by a classic model of plasma going through the diffusion region by Sweet and Parker \citep{sweet58, parker57}, it is conventional to think that 
plasma flows into the diffusion region from two upstream sides, where in the collisional case they collide and the bulk flows cannot interpenetrate. The two inflows meet at a stagnation point, 
are redirected by the magnetic tension force,
and exit the diffusion region on the downstream sides becoming the outflows (\autoref{fig:view}(b)). 
Computational results for reconnection in resistive magnetohydrodynamics (MHD) are 
quite consistent with the Sweet-Parker model \citep{biskamp86, scholer89, uzdensky2000}, even when generalized for asymmetric reconnection, i.e., between physically distinct inflows \citep{cassak07,birn08}.
With general acceptance of the logic of the Sweet-Parker model, this idea of how plasma flows through the diffusion region is commonly generalized to the case of collisionless reconnection as well \citep[see][]{daughton06}.


A refinement of the basic picture in \autoref{fig:view}(b) is to consider mixing of plasma from the two inflow regions, which is especially relevant for effectively collisionless reconnection. This is not included when treating the plasma from the two sides as a single fluid, but mixing effects have been found in particle-in-cell (PIC) simulations of collisionless reconnection, which treat ions and electrons using superparticles and can include kinetic effects that are not included by fluid models. 
Previous simulation studies have found mixing between two upstream populations, based on comparing PIC versus MHD simulation results \citep{malakit10} or particle trajectory and distribution function analyses \citep{zenitani16, hesse16, hesse17, dargent17}. 




In this work, we further investigate how collisionless plasma from the two inflow regions joins the outflows by labeling plasma particles from the two input regions in PIC simulations (\autoref{fig:view}(a)). We identify the large-scale flow patterns of ions and electrons from different source regions.  
Our findings are not consistent with flows turning the corner within the diffusion region on the same side (\autoref{fig:view}(b)) with some random mixing with the corresponding flow from the other region when joining the outflow. 
Instead, for both symmetric and asymmetric reconnection we demonstrate that the inflow from each side mostly crosses over to the other side, a new feature not yet reported by previous studies. 

\begin{figure*}[ht!]
\centering
\includegraphics[width=0.85\textwidth]{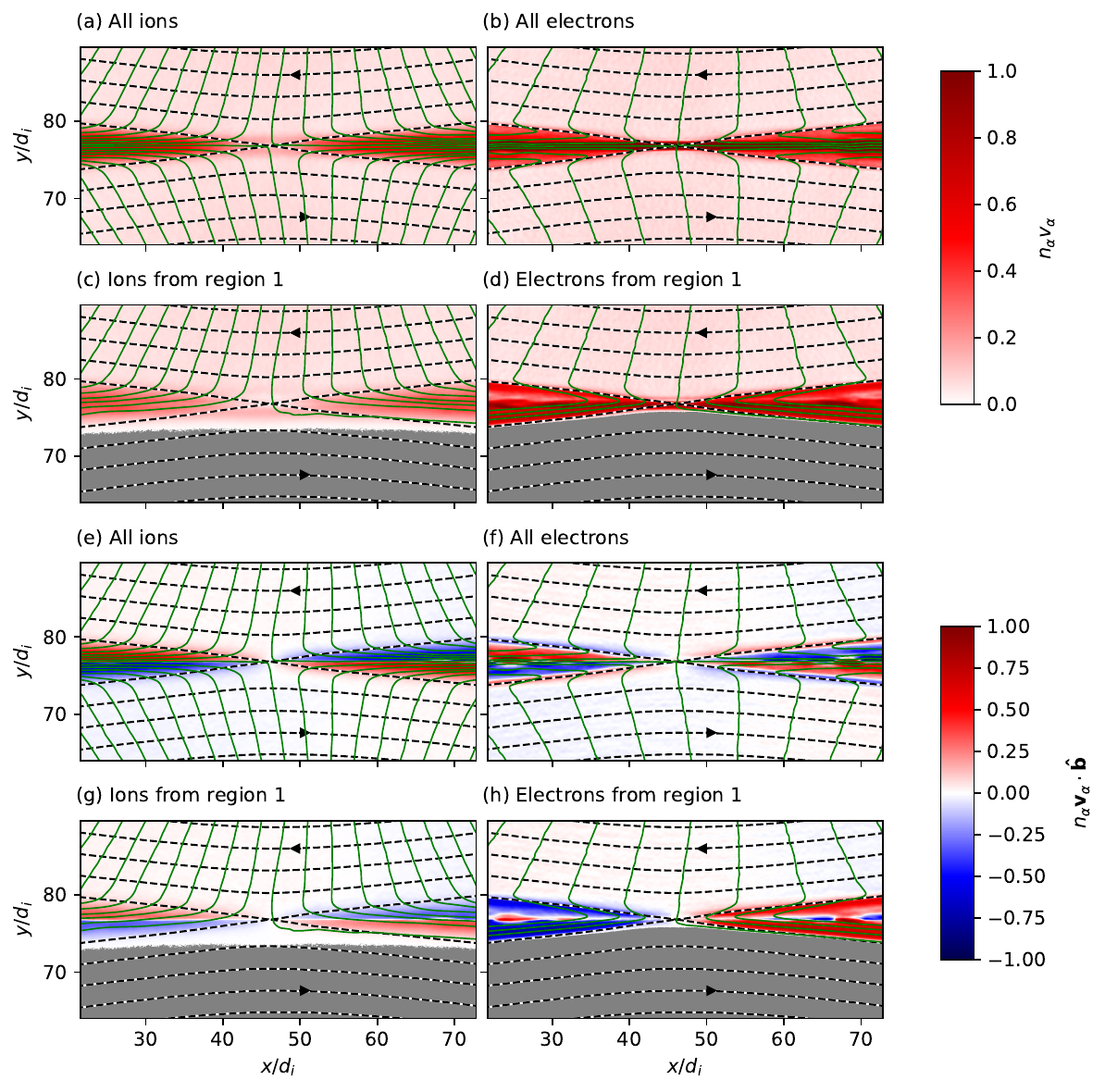}
\caption{Streamlines (green) of particle flux during symmetric collisionless magnetic reconnection overplotted on magnitude of particle number density flux for (a) all ions and (b) all electrons, and restricted to particles from the upper upstream region (region 1) as identified by particle labeling for (c) ions and (d) electrons. (e)-(h) Same species as in (a)-(d) but overplotted on the parallel number density flux. Grey areas indicate regions with particles per grid cell $\leq 1$. This figure demonstrates substantial flux crossover from the inflow region to the opposite side of the midplane. For electrons, we find a parallel flow toward the X-line followed by a crossover and parallel motion to join the outflow. For ions, there is a crossover in the diffusion region followed by parallel motion to join the outflow.}
\label{fig:nvdotb_sym}
\end{figure*}

\textit{Results}---
For the simplest case of symmetric reconnection without a guide field, \autoref{fig:nvdotb_sym}(a) shows streamlines (green) of ion flow {\it without} particle labeling, while \autoref{fig:nvdotb_sym}(c) shows the streamlines specifically for ions from region 1.  
From these plots, 
near the X-line 
we see that ions from one region stream across the midplane to the other region before joining the outflow, which we refer to as flow crossover.
The flow crossover is obscured when considering all ions without particle labeling in \autoref{fig:nvdotb_sym}(a).  
In that panel, ions streamlines do not cross the midplane while joining the outflow because the streamline calculation considers the ion distribution in terms of a single density and bulk flow, but from \autoref{fig:nvdotb_sym}(c) we see that in reality the two ion populations from each side are interpenetrating with a flow crossover.

\autoref{fig:nvdotb_sym}(b) and (d) are analogous plots for all electrons and electrons from region 1, respectively.  
The bulk flow of electrons differs from that of ions in that the electron flow first bends {\it toward} the X-line as it approaches a separatrix before reaching the midplane \citep{hoshino2001,wang10,lu10}.
For electrons from region 1 (\autoref{fig:nvdotb_sym}(d)), there is again a flow crossover to the other region before joining the outflow, but when considering all electrons together in \autoref{fig:nvdotb_sym}(b) this process is obscured: streamlines for the combined distribution remain on the same side as they join the outflow.

In \autoref{fig:nvdotb_sym}(a-d), the color scale indicates the magnitude of particle number density flux, while the same results are shown on the right, in \autoref{fig:nvdotb_sym}(e-h), with a color scale that instead indicates the number density flux parallel to $\mathbf{\hat b}$, the unit vector along the magnetic field.
The flow magnitude is weak in the inflow region and strong in the outflow region, as expected \citep{parker57}.
We see that in the outflows, the magnitude of the parallel number density flux is similar to the magnitude of the number density flux itself, implying that the flux is predominantly parallel to the magnetic field.

\begin{figure*}[hbtp]
\centering
\includegraphics[width=0.85\textwidth]{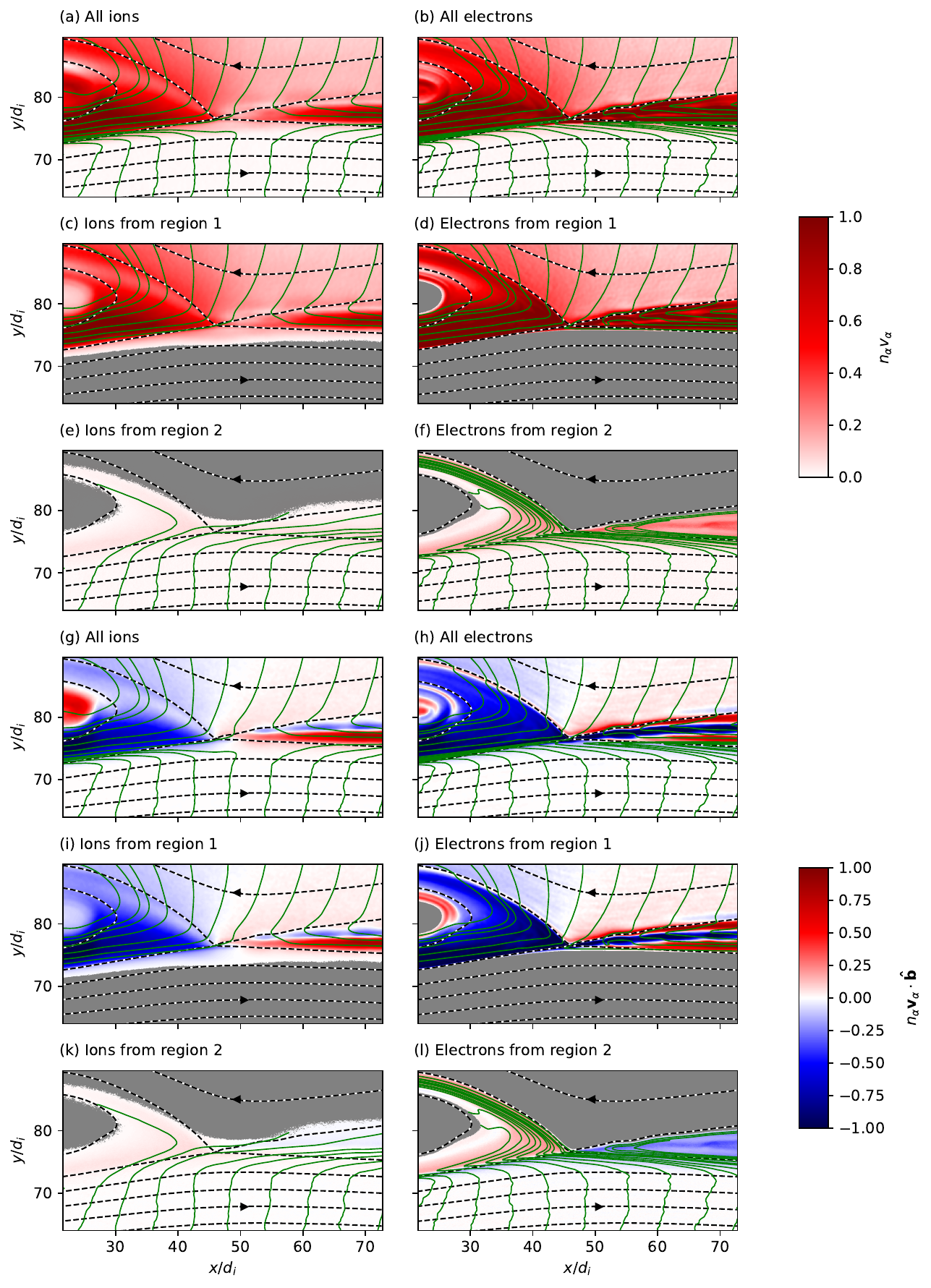}
\caption{
Like Figure 2, now showing that crossover features apply more generally to asymmetric reconnection with a guide field. }
\label{fig:nvdotb_asym}
\end{figure*}

In this sense, the physics of electron flow is more clearly demonstrated when labeling particles by source region, as opposed to the plot for all electrons (\autoref{fig:nvdotb_sym}(f)), in which the combined bulk electron flow crosses field lines in the outflow region.
The parallel flow of electrons from region 1 is positive on the right side and negative on the left (\autoref{fig:nvdotb_sym}(h)), corresponding to a parallel acceleration toward the X-line near their inflow separatrix \citep{hoshino2001, wang10,lu10,nan22}.
Then the electrons follow magnetic field lines toward the X-line, cross the midplane, and join the outflow on the other side, maintaining the same sign of parallel flow.
(During this process, the electrons can undergo reflection and Fermi acceleration as noted by \citet{drake06}.)
However, \autoref{fig:nvdotb_sym}(f) for all electrons combined indicates a second kick {\it away} from the X-line immediately after the separatrix, with the sign of parallel flow suddenly reversing.
We can now see that the combined parallel electron flow in the outflow regions (across the separatrices) is dominated by electrons from the {\it opposite} side, which have the opposite parallel flow direction due to their acceleration toward the X-line in {\it their} inflow region.
The apparent second kick in the combined electron flow at each separatrix is actually a result of 
including electrons that have crossed over from the opposite side.

For asymmetric reconnection, with more complex flows, we find that the crossover pattern still persists, demonstrating generality of the feature. In our guide-field asymmetric reconnection simulation, the asymptotic density ratio of upper inflow side (side 1) over lower inflow side (side 2) is 10 and the asymptotic reconnection magnetic field ratio is 0.5.  We see that \autoref{fig:nvdotb_asym}(a) and (b) are not very different from \autoref{fig:nvdotb_asym}(c) and (d), meaning the plasma from the higher density side (side 1) dominates the flow in the outflow region.  This is expected based on prior results for asymmetric reconnection \citep[e.g.,][]{malakit10}.  
Note also that the total flux is not exactly the sum of fluxes from regions 1 and 2, because some region-3 plasma remains in the outflows, especially in the part of the left outflow region that is colored grey (indicating $\leq1$ particle per grid cell) or white for region 1 in \autoref{fig:nvdotb_asym}(c-d) and (i-j).

The flow crossover that we identify from PIC simulations could be verified by {\it in situ} observations of asymmetric reconnection in a collisionless plasma.
Being asymmetric implies that the two inflow plasmas have distinct physical properties, such as density, temperature, and composition, for example at a planetary magnetopause where the solar wind encounters a magnetosphere.
Thus there is an overall gradient in a given quantity across the reconnection zone.  
However, because of the flow crossover reported here, 
we predict that the bulk flow on one side of the initial outflow should exhibit properties characteristic of plasma from the {\it other} inflow side.
The crossover should manifest in a {\it reversed gradient} in plasma properties 
in the initial outflow, i.e., a local gradient in the opposite direction from the overall gradient across the reconnection zone.
Further work could search for this effect in {\it in situ} observations, and additional simulation work could comprehensively model reversed gradients.

\textit{Conclusions}---
By labeling particles in particle-in-cell simulations according to their initial inflow region, we reveal a bulk flow crossover in collisionless magnetic reconnection, in which plasma flows across the midplane to the other side  before joining an outflow  (\autoref{fig:view}). 
We attribute the crossover feature to generation of parallel flow.  
Ion parallel flow is mostly generated within the ion diffusion region, while electron parallel flow is mostly created outside the electron diffusion region, due to acceleration toward the X-line as it crosses a separatrix, followed by parallel motion along the magnetic field and across the midplane to join the outflow. 
Importantly, the flow crossover and parallel flow generation is sufficiently dominant that the reconnection outflow is more parallel than perpendicular to the magnetic field.  The flow crossover can be found not only in simple anti-parallel symmetric reconnection but also in more complicated guide-field asymmetric reconnection (\autoref{fig:nvdotb_asym}), suggesting it is a general feature of collisionless reconnection. 
This flow crossover could be identified in collisionless space plasmas by {\it in situ} measurements, because plasma from one inflow side can affect or even dominate plasma properties in the outflow on the other side
(e.g., compare \autoref{fig:nvdotb_sym}(f) and (h)).  
We predict that in the outflow sufficiently close to an asymmetric reconnection site, the plasma properties exhibit a locally reversed gradient, in a direction opposite to the large-scale gradient across the reconnection zone.


This research made use of computing resources kindly provided by the US National Energy Research Scientific Computing Center during the initial phases.
Support was provided by grants RTA5980003 and MRG6180176 from the Thailand Research Fund and RTA6280002 from Thailand Science Research and Innovation, the Kasetsart University Research and Development Institute (KURDI), and grant RGNS 64–027 by the Office of the Permanent Secretary, Ministry of Higher Education, Science, Research and Innovation (OPS MHESI), Thailand Science Research and Innovation (TSRI) and Kasetsart University. This research was also partially supported by the National Science and Technology Development Agency (NSTDA) and National Research Council (NRCT): High-Potential Research Team Grant Program (N42A650868), 
and by NASA grant 80NSSC20K1813.
MAS and PAC gratefully acknowledge the Thai NSRF via the Program Management Unit for Human Resources \& Institutional Development, Research and Innovation (B39G670013) for support for a research meeting with the coauthors in Thailand.  
PAC acknowledges DOE grant DE-SC0020294 and NASA grant 80NSSC24K0172.

\appendix

\section{Simulation Setup and Parameters}




The simulations of collisionless magnetic reconnection are performed using \texttt{p3d}, a 2.5D fully kinetic particle-in-cell (PIC) code \citep{zeiler02}.  The results obtained from \texttt{p3d} are given in normalized units. The magnetic field $B_0$ and plasma density $n_0$ represent the reference magnetic field strength and density. Lengths are normalized to the reference ion inertial scale $d_{i0} = c \sqrt{m_i/ 4\pi n_0 e^2}$, where $m_i$ is the ion mass, $e$ is the ion charge, and $c$ is the speed of light. Times are normalized to the reference ion cyclotron time $\omega^{-1}_{ci0} = m_i c / (e B_0)$. Velocities are normalized to the Alfv\'en speed based on the reference magnetic field strength and density, $c_{A0} = B_0 / \sqrt{4 \pi m_i n_0}$. 
Temperatures are normalized to $T_0=m_i c^2_{A0}$. The speed of light $c$ is set to $15c_{A0}$. 
The electron-to-ion mass ratio is $m_e/m_i = 1/25$.

The simulations are performed in a rectangular domain of size $L_x \times L_y = 204.8 d_{i0} \times 102.4 d_{i0}$. The boundary conditions are periodic in all directions. The initial conditions for symmetric reconnection are a double Harris sheet. The initial condition for asymmetric reconnection is a double current sheet with the magnetic field and temperature varying across current sheets from upstream side 1 to upstream side 2 as tanh functions for width $w_0$. The density varies in such a way as to fulfill the total pressure balance \citep[for more details, see][]{malakit10}.
The initial velocity distribution function is Maxwellian.
Near the current sheet, the distribution function is shifted to a bulk velocity in opposite directions for ions and electrons in proportion to the temperature of each species so as to provide a current density that satisfies Amp\`ere's law.
The parameters used in the simulations are shown in \autoref{tab:input}.  A small perturbation is used to initiate reconnection on the upper left ($x=0.25 L_x, y=0.75L_y$) and lower right ($x=0.75L_x, y=0.25L_y$) of the simulation box. 
The particles per grid cell for the reference density ($ppg$) is equal to 100.

The ions and electrons in the simulations are labeled according to their initial regions, allowing us to categorize each species of plasma into 3 populations (see \autoref{fig:view}(a)).  Region-1 and region-2 populations refer to particles that are initially in the upstream regions with magnetic field pointing to the $-x$ and $+x$ directions, respectively.  The region-3 population comprises particles in the transition region between regions 1 and 2, within which the initial current sheet resides. In our simulations, we set the thickness of this transition region to be $5.0 d_{i0}$.  With the particle-labeling technique, we are able to evaluate bulk quantities such as density, bulk velocity, and pressure of plasma of each population separately. 

Reconnection in the simulations is allowed to grow until the reconnection rate becomes relatively steady. In addition, we wait until the region-3 population is pushed far downstream, so the effects of the initial conditions are minimized. The data are averaged over 200 time slices over a total period of 1 $\omega^{-1}_{ci0}$ 
at the end of the simulations to reduce noise. While our simulation box has two reconnection regions, we use the upper-left reconnection region for further analysis because the reconnection there happens to be cleaner, meaning there are no secondary islands or other complications.


\begin{table}[]
    \centering
    \begin{tabular}{l c c}
        \hline \hline
        Parameter &  Symmetric  & Asymmetric \\
        \hline
        Simulation $x$-size $L_x$ & 204.8  & 204.8  \\
        Simulation $y$-size $L_y$ &  102.4  & 102.4  \\
        Grid scale $\Delta x$, $\Delta y$ & 0.05  & 0.05  \\
        Time step $\Delta t$ & 0.005  & 0.005  \\
        Initial current sheet width $w_0$ & 1.0  & 1.0  \\
        Region 1 magnetic field $B_1$ &  1.0  & 1.0  \\
        Region 2 magnetic field $B_2$ &  1.0  & 2.0  \\
        Region 1 plasma density &  1.0  & 1.0  \\
        Region 2 plasma density &  1.0  & 0.1  \\
        Region 1 ion temperature & 0.25  & 1.33 \\
        Region 2 ion temperature & 0.25  & 3.33 \\
        Region 1 electron temperature & 0.25  & 0.667 \\
        Region 2 electron temperature & 0.25  & 1.667 \\
        Guide magnetic field  & 0 & 1\\
        \hline \hline
    \end{tabular}
    \caption{Parameters of the 2.5D PIC simulation runs used in this study for symmetric reconnection with no guide field and for asymmetric reconnection with a guide field.}
    \label{tab:input}
\end{table}

\section{Role of Parallel Flow}
The parallel flow is significant for understanding the crossover.
The midplane is where $B_x=0$, and in the case of zero guide field $B_z$,  the magnetic field is generally in the $y$-direction, which implies that flow across the midplane must be parallel flow. 
Flow that is strictly perpendicular to the magnetic field, such as the $\mathbf{E\times B}$ drift for frozen-in flow, is not able to cross over the midplane.
Indeed, even with a guide field, the $\mathbf{E\times B}$ drift affects all plasma components in the same way and cannot be the reason for one flow to cross paths with another flow.
For ions, \autoref{fig:nvdotb_sym}(g) shows that the flow crosses the midplane in the ion diffusion region near the X-line, which relates to 
their finite Larmor radius and violation of the frozen-in condition, and after the crossover in the ion diffusion region there is a parallel outflow.
For electrons, the inflow sharply changes its direction toward the X-line shortly before the flow crosses the separatrix \citep{lu10}.  
Then after the electron bulk flow from one region crosses the separatrix, it basically follows a magnetic field line across the midplane (i.e., the parallel flow retains the same sign) to join the outflow on the opposite side \autoref{fig:nvdotb_sym}(h).

In asymmetric reconnection, we emphasize that for plasma from the higher density side to dominate the flow in the outflow region, even on the other side of the midplane (where $B_x=0$), the plasma from the higher density side 
experienced mechanisms that generate parallel flow, helping the plasma to cross the midplane. Also, the plasma from the lower density inflow (region 2) has a small contribution to the total flow, yet it is not constrained to its own side as suggested by the streamlines of the total flow \autoref{fig:nvdotb_asym} (a) and (b). We find that once the plasma flows into the area around the reconnection site, it again undergoes flow-parallelization mechanisms, which consequently lead to the flow crossover of plasma from the lower density side (\autoref{fig:nvdotb_asym} (e-f)). As in the case of symmetric reconnection, the majority of the particle flux in the outflow is parallel to the magnetic field rather than perpendicular. We also note that adding complexity to the reconnection can lead to complex crossover patterns.  For example, we observe left-right asymmetry in the crossover flow. Nevertheless, ions and electrons from one inflow region consistently tend to end up on the other side of the outflow (\autoref{fig:nvdotb_asym} (e-f)). 

A previous simulation study by \citet{nan22} provided a detailed analysis of the electron acceleration along the separatrices toward the X-line, including a contribution related to the parallel electric field, and the pressure gradient force has also been shown to contribute to the parallel acceleration of electron flow \citep[see also][]{hesse17}.
Parallel acceleration of electron flow due to parallel electric fields has indeed been observed by the Magnetospheric Multiscale (MMS) mission in association with magnetic reconnection in Earth's magnetosphere \citep{eriksson18, norgren20}.



In the standard picture of symmetric reconnection in \autoref{fig:view}(b), the flow needs to stagnate at the X-line (labeled by ``S'' in the figure), and for asymmetric reconnection in this picture there is still a single stagnation point, which is displaced from the X-line \citep{cassak07}. 
However, in our simulations of collisionless reconnection with particle labeling, 
after electrons and ions approach the X-line, most of them cross the current sheet, 
with distinct stagnation points (labeled ``S1'' and ``S2'' in 
\autoref{fig:view}(c)) located on the opposite side from the initial inflow.
Farther from the X-line, some of the bulk flow does not pass through the diffusion region and does not cross over (indicated by dashed lines in the Figure), and is instead carried along to the outflow on the same side of the midplane according to the $\mathbf{E\times B}$ drift.

Note that the picture of plasma flow from one inflow side mostly not crossing the midplane (\autoref{fig:view}(b)) is reasonable if the plasma is collisionally thickened. The collisions between plasmas from the two inflow sides can prevent interpenetration of each plasma flow originally from one side into the other.  For strongly collisional flows, the only mechanism that allows plasma to penetrate to the other side of an interface is collisional diffusion, which is a slow process. 

Our results based on particle labeling also provide insight into some previous results in the literature.  
For example, \citet{hesse17} discussed multiple source populations in the outflow regions (without the use of particle labeling) and referred to a process of ``mixing'' 
\citep[see also][]{shay98,zenitani16, bessho16}.
However, while the term ``mixing'' suggests a random or thermodynamic process, our results indicate a {\it systematic} crossover of the bulk flow.  
In particular, for the symmetric reconnection case the sign of the parallel outflow of electrons on each side of the midplane indicates a clear domination by electrons from the {\it other} side (comparing \autoref{fig:nvdotb_sym}(f) and (h)). 

\bibliography{ref}

\end{document}